\documentclass[11pt]{article}
\usepackage{xspace}
\usepackage{graphicx}
\usepackage{amsmath}
\usepackage{amssymb}
\usepackage{color}

\definecolor{Red}{rgb}{1,0,0}
\definecolor{Green}{rgb}{0,1,0}
\definecolor{Blue}{rgb}{0,0,1}
\definecolor{Black}{rgb}{0,0,0}

\def\beq{\begin{equation}}
\def\eeq#1{\label{#1}\end{equation}}
\def\eeqn{\end{equation}}

\def\beqa{\begin{eqnarray}}
\def\eeqa#1{\label{#1}\end{eqnarray}}
\def\eeqan{\end{eqnarray}}

\let\bar=\overbar

\def\Dslash{\not{\hbox{\kern-4pt $D$}}}
\def\dslash{\not{\hbox{\kern-2pt $\del$}}}

\def\msb{{\bar{\ssstyle M \kern -1pt S}}}

\def\Title#1{\begin{center} {\Large {\bf #1} } \end{center}}

\begin{document}

\Title{Directional Dark Matter Search and Velocity Distribution}

\bigskip\bigskip


\begin{raggedright}  

Keiko I. Nagao \footnote{The talk is based on the collaboration with Tatsuhiro Naka (Nagoya University) and Mihoko. M. Nojiri (KEK and Kavli IPMU)}\\
{\it National Institute of Technology, Niihama College}\\

\bigskip
\end{raggedright}

\begin{center}
\subsection*{Abstruct}
\end{center}
Directional detection of dark matter is the next generation experiment, which is expected to have better  back ground rejection efficiency than conventional direct search. 
Another intriguing possibility of the experiment by means of the directional information is measurement the velocity distribution of dark matter. Especially, it will be potent to figure out whether the velocity distribution is anisotropic. 
Supposing three distribution models, we discuss the possibility in one of the directional dark matter searches, nuclear emulsion detector.
\\
{\small
\begin{flushleft}
\emph{To appear in the proceedings of the Interplay between Particle and Astroparticle Physics workshop, 18 -- 22 August, 2014, held at Queen Mary University of London, UK.}
\end{flushleft}
}

\section{Introduction}
Cosmological observations, including the cosmic microwave background, baryon acoustic oscillations,  and supernovae, proves that dark matter makes up about 23\% of the energy density of the Universe. Many studies to reach its nature have been performed from both theoretical and experimental sides. Especially, the weakly interacting massive particle (WIMP) is a promising candidate of dark matter in the theoretical models. WIMP with the electroweak-scale mass is known to satisfy the constraint of the relic abundance in a natural way, it is called as WIMP miracle.

Three types of experiment, direct and indirect detections, and colliders, are hopeful to detect the WIMP. In the direct detection, dark matter can scatter a target nucleus in underground facility, and we can detect dark matter by measuring its recoil energy. In the last few decades many collaborations have given constraints for the mass and interaction of dark matter. Currently, WIMP-nucleon interaction cross section larger than about $10^{45}$ cm$^2$ has been excluded with O(100) GeV mass by LUX collaboration \cite{Akerib:2013tjd}, and other collaborations also support the result. 

Since dark matter rarely interacts with other particles, back ground rejection is a major problem in the direct search.
Detection with directional sensitivity, which can solve the problem, is the next generation project for direct detection of dark matter. Due to the revolution around the center of the Milky Way Galaxy, the Solar System receives dark matter flow steadily, that is so-called dark matter wind. In other words, the dark matter signals are expected to come from the direction of the wind. Thus the directional information is powerful clue to distinguish dark matter and back ground signals in the direct detection experiment. The dark matter flow is also affected by yearly and daily round of the Earth. Though such motions can be observed in the ordinary direct detection through the variation of the signal number, 
 detection with directional sensitivity makes it possible to confirm more certainly. Several projects of directional direct detection have launched. Most of them are gasses detectors which aim to detect dark matter through spin-dependent interaction \cite{Santos:2013hpa}, on the other hand, a project uses solid target that is suitable for spin-independent interaction. The latter one is the nuclear emulsion detector \cite{D'Ambrosio:2014wma}, in which the recoil targets leave tracks inside the emulsion.

In the paper, we study the other possibility of the directional dark matter search. Directional sensitivity can open the way for distribution of dark matter, especially, it is potent to investigate whether the distribute is specially symmetric or not. Adopting some sample distribution models, we simulate the angular distribution of dark matter signal in the nuclear emulsion detection for each model. 

\section{Directional Dark Matter Search in Nuclear Emulsion Detector}
\subsection{Direct Detection}
In the direct detection experiment, the constraint for the interaction cross section of dark matter is derived from the event rate $R$, which is conventionally expressed in unit of /counts/kg/day/keV.
The differential event rate depend on the local energy density of dark matter $\rho_0$, WIMP mass $m_\chi$, the mass of target nuclei $m_N$, WIMP velocity $v$, and the velocity distribution of dark matter in the Solar System $f(v)$ :
\begin{eqnarray}
\frac{dR}{dE_R}=\frac{\rho_0}{m_\mathrm{N}m_\chi} \int_{v_\mathrm{min}}^{v_\mathrm{max}} \frac{1}{v}\  f({v}) \frac{d\sigma(v, E_R)}{dE_{R}} \ d^3v.
\label{eq: dRdE}
\end{eqnarray}
The maximal velocity $v_\mathrm{max}$ is the local Galactic escape velocity, and the minimum velocity related to the recoil energy as $v_\mathrm{min}=\sqrt{m_NE_R/2 m_\chi^2}$.

\subsection{Nuclear Emulsion Detector}
\begin{table}[t!]
\begin{center}
\begin{tabular}{|c||r||}
\hline
\multicolumn{1}{|c|}{\,} & \multicolumn{1}{c|}{Weight(\%)}  \\ \hline\hline
Ag                                & 39.65                                           \\ \hline
Br                                 & 29.01                                           \\ \hline
O                                  & 11.76                                          \\ \hline
C                                  & 11.72                                            \\ \hline
N                                  & 4.57                                               \\ \hline
\end{tabular}
\caption{Composition of nuclear emulsion used in the directional dark matter detection.}
\end{center}
\label{tab:composition}
\end{table}
Direct detection using the nuclear emulsion is one of the directional direct search experiments. Nuclear emulsion is a kind of photographic films. Once the nuclear emulsion is exposed by a charged particle, its track appears in the nuclear emulsion after development treatment. It can be utilized to detect an electrically neutral particle through its decay or scattering involving charged particles. It has been used for neutrino detection in OPERA collaboration, and is also applicable to dark matter detection. Especially, it can be the detector with directional sensitivity since the emulsion can record tracks of recoiled particles, which is related to the direction of incoming WIMP.

\subsubsection*{Target Atoms}
Nuclear emulsion is not only a detector but also the target material itself. It consists of silver bromide (AgBr) crystal, which is the basis of photographic film, and gelatin. The gelatin contains oxygen, calcium, and nitrogen.  To summarize, there are relatively heavy target atoms: Ag and Br, and light atoms: O, C, and N in nuclear emulsion used in the directional direct detection. Its composition with the weight ratio is shown in Table 1.

\subsubsection*{Resolution of the nuclear emulsion detector}
High resolution detector is required to measure the slight difference of the velocity distribution. The angular resolution of nuclear emulsion used in the direct detection is 15-20$^\circ$, and the spatial resolution is 70 $\sim$100 nm. The spatial resolution determines the detectable recoil energy for each target atom; 220 keV for Ag, 160 keV for Br and 33 keV for light target atoms. Afterward, these constraints for the recoil energy is applied in the numerical calculation. 

\section{Numerical Calculation}
\subsubsection*{Velocity Distribution Models}
In the numerical calculation, we simulated WIMP scattering supposing three models of WIMP velocity distribution, which is listed below. \\

\noindent (a) Maxwell Distribution\\
The differential event rate is affected not only by the interaction rate of dark matter but also by its velocity distribution as implied in Eq.(\ref{eq: dRdE}).
Conventionally, spatially symmetric distributions is adopted as the distribution models. The most popular expression is the Maxwell distribution,
\begin{eqnarray}
f({\bf v})=\frac{1}{\left( \pi v_0^2 \right)^{3/2}}\ e^{-({\bf v}+{\bf v}_E)^2/v_0^2}.
\end{eqnarray}\\

\noindent (b) Tidal Stream\\
However, irregular distributions are also suggested by the numerical simulations and observations. For instance, a N-body simulation shows that tidal streams can be caused due to flow of subhalos falling into the Milky Way Galaxy \cite{Ling:2009eh}. The velocity distribution is distorted, and the tangential component of the velocity to the motion of the Solar System becomes non-symmetric. In that case, the number of the WIMP signals from the direction of WIMP wind is expected to decrease compared to the case of Maxwell distribution. Such tidal streams are observed in the Sagittarius dwarf galaxy \cite{Ivezic:2000ua}.\\

\noindent (c) Debris Flow\\
The other irregular distribution we use in the numerical calculation is affected by so-called debris flow \cite{Kuhlen:2012fz}. For this case, the velocity distribution has a  substructure as well as the tidal streams, however, materials recently falls into the Milky Way Galaxy from subhalos and not completely phase-mixed.

\subsubsection*{Numerical Result}
In Figure 1-3, the angular distribution of the WIMP signals in the directional direct detection with nuclear emulsion, i.e., the track length of the recoil nuclei in the nuclear emulsion detector, is shown. The horizontal and the vertical axes represent $\cos{\theta}$ and the track length, respectively, where $\theta$ is scattering angle of the recoil nucleus to the direction of dark matter wind. Distribution for Maxwell distribution, 
the velocity distribution with tidal streams \cite{Ling:2009eh}, and with debris flow \cite{Kuhlen:2012fz} correspond to Figure 1, 2, and 3, respectively.  The elastic scattering is supposed in the Monte Carlo simulation for all the figures. 
The track length is derived from the recoil energy of nuclei, however, it depends on the type of target atoms \cite{Nagao:2012gp}. Target atoms are Ag, Br, O, C, and N, and the composition is same as the realistic nuclear emulsion as shown in Table 1. Left panels are for $m_\chi = 200$ GeV, and right panels are for $m_\chi= 800$ GeV. The number of signals in small $\cos{\theta}$ region vanishes for the velocity distribution with tidal streams and debris flow, though some signals remain for the Maxwell distribution. The difference is prominent in the case of large WIMP mass, since the track length is extended due to large recoil energy.
\begin{figure}[h!]
\hspace{-2.2cm}
 \begin{minipage}{0.7\hsize}
  \begin{center}
   \includegraphics[width=70mm]{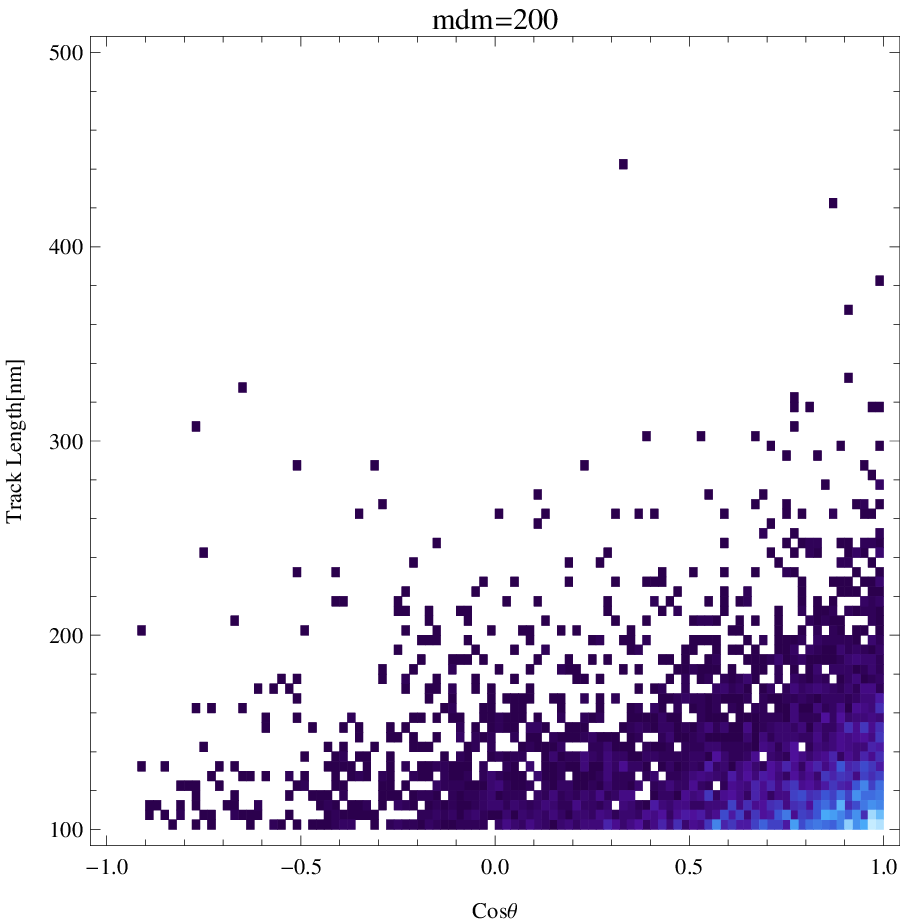}
     \includegraphics[width=6mm]{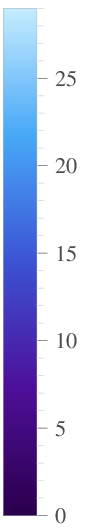}
  \end{center}
 \end{minipage}
 \hspace{-2.5cm}
 \begin{minipage}{0.7\hsize}
 \begin{center}
  \includegraphics[width=70mm]{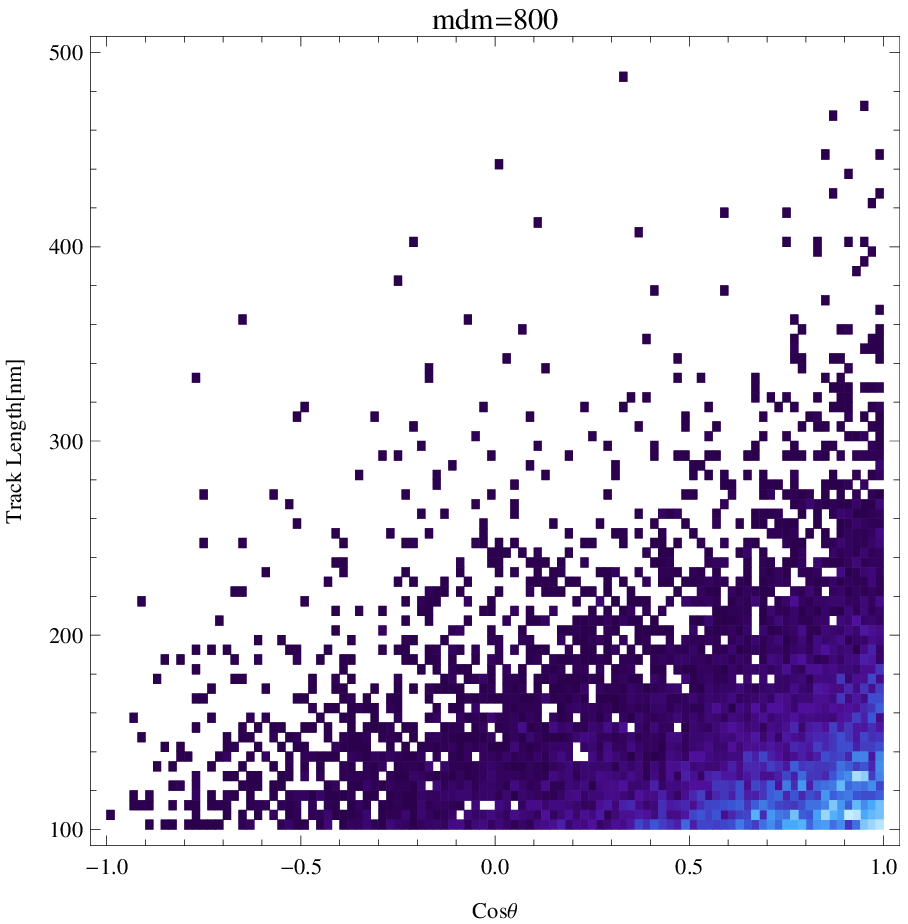}
   \includegraphics[width=6mm]{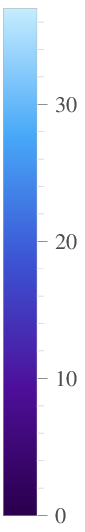}
 \end{center}
 \end{minipage}
   \label{fig:maxwell}
 \caption{Angular distribution of WIMP signal for (a) Maxwell distribution. Left: $m_\chi=200$ GeV, Right: $m_\chi=800$ GeV. The horizontal and the vertical axes represent $\cos{\theta}$ and the track length of WIMP signal in the nuclear emulsion detector, respectively. }
 \vspace{1cm}
  \hspace{-2.2cm}
 \begin{minipage}{0.7\hsize}
  \begin{center}
   \includegraphics[width=70mm]{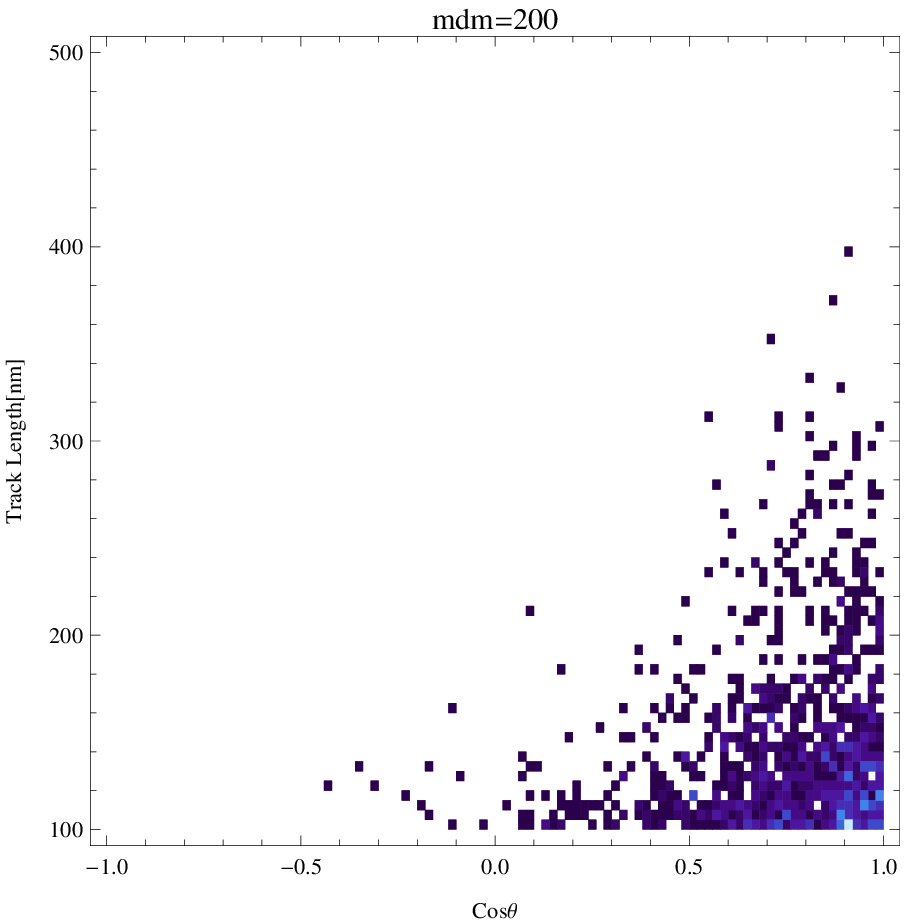}
     \includegraphics[width=6mm]{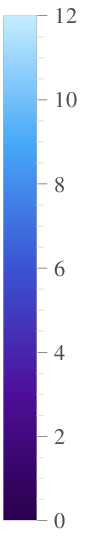}
  \end{center}
 \end{minipage}
 \hspace{-2.5cm}
 \begin{minipage}{0.7\hsize}
 \begin{center}
  \includegraphics[width=70mm]{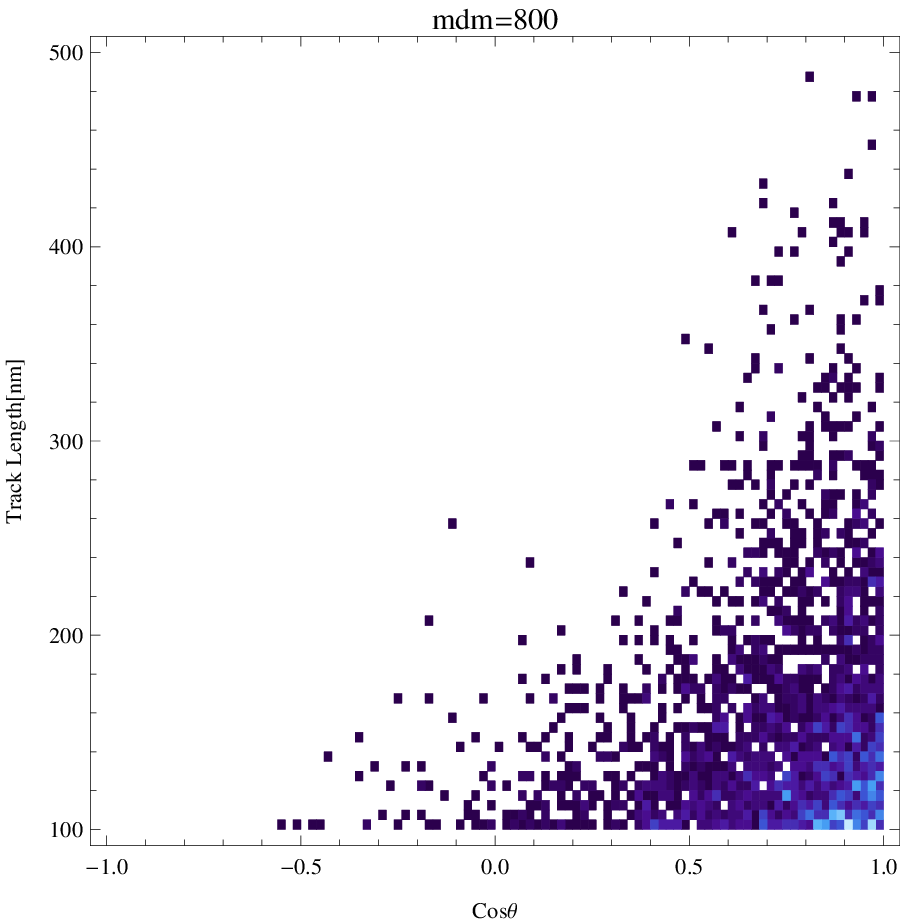}
   \includegraphics[width=6mm]{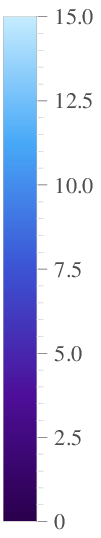}
 \end{center}
 \end{minipage}
 \label{fig:LNAT}
  \caption{Legend is same as Figure 1 but for (b) the distribution with tidal stream.}
 \end{figure}
 \begin{figure}[ht]
  \hspace{-2.2cm}
  \begin{minipage}{0.7\hsize}
  \vspace{1cm}
  \begin{center}
   \includegraphics[width=70mm]{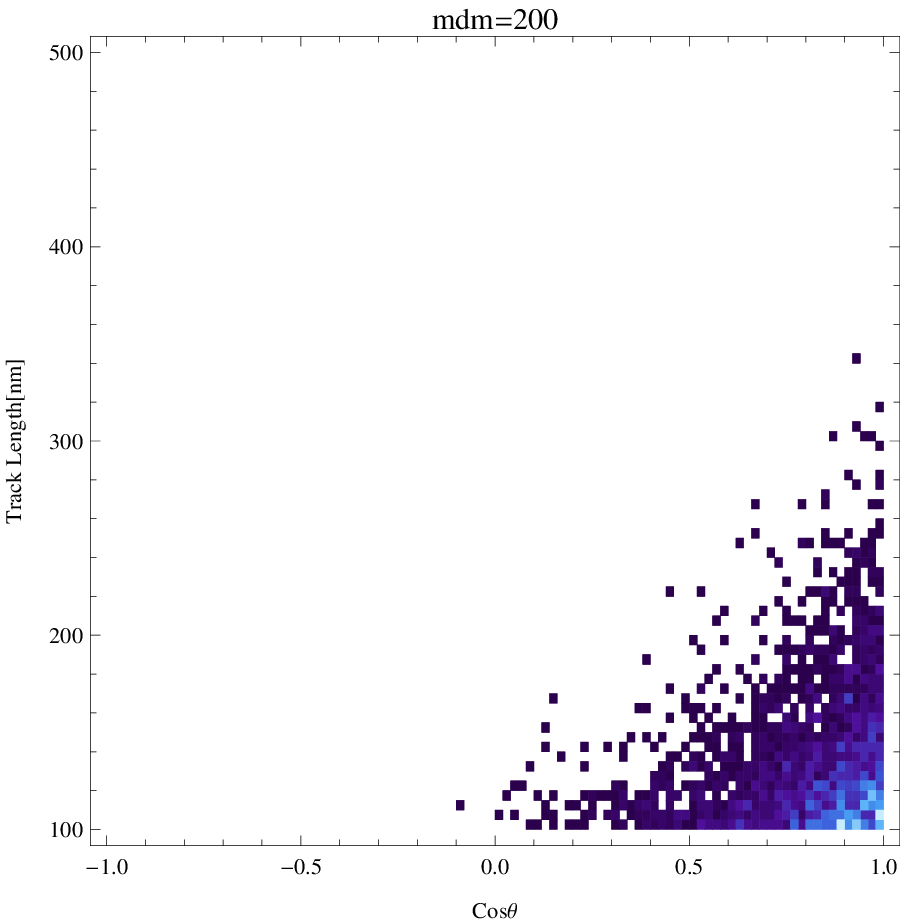}
     \includegraphics[width=6mm]{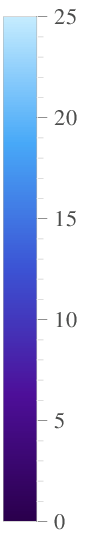}
  \end{center}
 \end{minipage}
 \hspace{-2.5cm}
 \begin{minipage}{0.7\hsize}
 \begin{center}
   \vspace{1cm}
  \includegraphics[width=70mm]{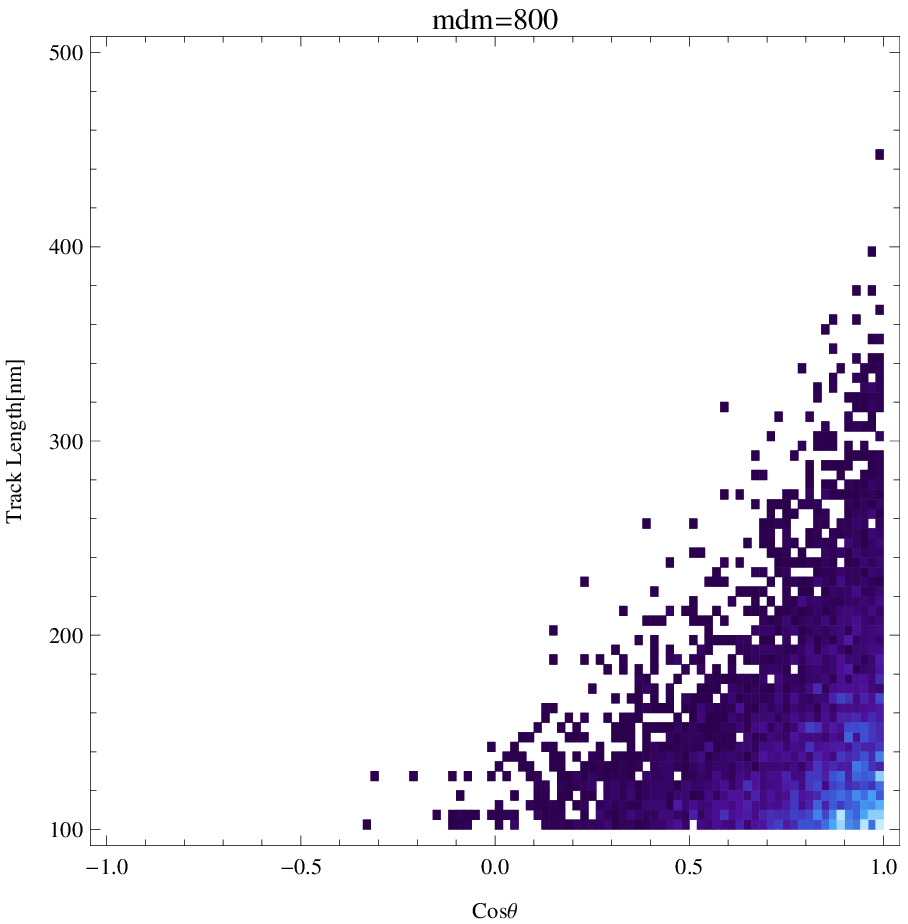}
   \includegraphics[width=6mm]{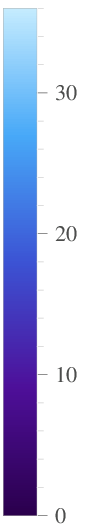}
 \end{center}
 \end{minipage}
  \begin{center}
  \label{fig:KLS}
 \caption{Legend is same as Figure 1 but for (c) the velocity distribution with debris flow.}
\end{center}
\end{figure}

\section{Summary}
The possibility to measure the velocity distribution of dark matter in the directional detection is discussed. By means of the directional information, it will be notably useful to determine whether the velocity distribution is anisotropic or not. We focused a standard Maxwell distribution, and two distributions which have components falling into the Milky Way Galaxy from subhalos in the paper.
Angular distribution of the WIMP signal for three distribution models are simulated and compared. In the simulation, we suppose the detection by  the nuclear emulsion, which is one of the directional detectors. It is distinguishing that signals remain in small $\cos{\theta}$ region for Maxwell distribution while they vanishes for the other two models. The behavior can be observed even in the region where the recoil energy is higher than the threshold of the detector.

\end{document}